\documentclass[aps,prd,nofootinbib,twocolumn,superscriptaddress,preprintnumbers]{revtex4}

\usepackage{amssymb}
\usepackage{amsmath}
\usepackage{epsfig}
\usepackage{hyperref}
\usepackage{breakurl}
\usepackage{amsbsy}

\usepackage{undertilde}

\makeatletter
\def\simgt{\mathrel{\lower2.5pt\vbox{\lineskip=0pt\baselineskip=0pt
           \hbox{$>$}\hbox{$\sim$}}}}
\def\simlt{\mathrel{\lower2.5pt\vbox{\lineskip=0pt\baselineskip=0pt
           \hbox{$<$}\hbox{$\sim$}}}}
\makeatother

\newcommand{\be}{\begin{equation}}
\newcommand{\ee}{\end{equation}}
\newcommand{\bea}{\begin{eqnarray}}
\newcommand{\eea}{\end{eqnarray}}
\newcommand{\Eq}[1]{Eq.~(\ref{#1})}
\newcommand{\Eqs}[2]{Eqs.~(\ref{#1}) and (\ref{#2})}
\newcommand{\Sec}[1]{Sec.~\ref{#1}}
\newcommand{\App}[1]{App.~\ref{#1}}
\newcommand{\Ref}[1]{Ref.~\cite{#1}}

\renewcommand{\Im}{\mathop{\text{Im}}}
\newcommand{\Arg}{\mathop{\text{Arg}}}

\newcommand{\hc}{\text{h.c.}}

\newcommand{\vev}[1]{\langle #1 \rangle}

\newcommand{\MPl}{M_{\rm Pl}}
\newcommand{\bPhi}{{\boldsymbol \Phi}}

\newcommand{\bW}{{\boldsymbol W}}

\newcommand{\bX}{{\boldsymbol X}}

\newcommand{\bK}{{\boldsymbol K}}
\newcommand{\bZ}{{\boldsymbol Z}}
\newcommand{\bT}{{\boldsymbol T}}
\newcommand{\bP}{{\boldsymbol P}}

\newcommand{\bs}[1]{\boldsymbol #1}

\begin{document}

\preprint{MIT-CTP 4243}

\title{Supergravity Computations without Gravity Complications}

\author{Clifford Cheung}
\affiliation{Berkeley Center for Theoretical Physics, 
  University of California, Berkeley, CA 94720, USA}
\affiliation{Theoretical Physics Group, 
  Lawrence Berkeley National Laboratory, Berkeley, CA 94720, USA}

\author{Francesco D'Eramo}
\affiliation{Center for Theoretical Physics, 
  Massachusetts Institute of Technology, Cambridge, MA 02139, USA}
  
\author{Jesse Thaler}
\affiliation{Center for Theoretical Physics, 
  Massachusetts Institute of Technology, Cambridge, MA 02139, USA}

\begin{abstract}

The  conformal compensator formalism is a  convenient and versatile representation of  supergravity (SUGRA) obtained by gauge fixing conformal SUGRA.  Unfortunately, practical calculations often require cumbersome manipulations of component field terms involving the full gravity multiplet.  In this paper, we derive an alternative gauge fixing for conformal SUGRA which decouples these gravity complications from SUGRA computations.  This yields a simplified tree-level action for the matter fields in SUGRA which can be expressed  compactly in terms of superfields and a modified conformal compensator.  Phenomenologically relevant quantities such as the scalar potential and  fermion mass matrix are then straightforwardly obtained by expanding the action in superspace.

\end{abstract}

\maketitle

\section{Introduction}

Supersymmetry (SUSY) is a well-studied and highly-motivated extension of the standard model.  While some aspects of SUSY phenomenology may be understood purely in the limit of global SUSY, others require 
the full machinery of supergravity (SUGRA).  For example, SUGRA plays an essential role in the super-Higgs mechanism, whereby the goldstino of spontaneous SUSY breaking is eaten to become the longitudinal mode of the gravitino \cite{Wess:1992cp,Cremmer:1978hn}.  This induces a non-zero mass for the gravitino $m_{3/2}$, which plays a crucial role in SUSY cosmology and collider phenomenology \cite{Martin:1997ns,Moroi:1993mb}.  Furthermore, the full  SUGRA formalism is required for a proper description of ``no-scale'' SUSY breaking \cite{Wess:1992cp,Lahanas:1986uc,Luty:2005sn}, which arises when moduli mix directly with the gravity multiplet.   

Despite its clear phenomenological significance,
not all of the SUGRA formalism is actually relevant for practical calculations.  For instance, couplings to the graviton are more or less unimportant for SUSY phenomenology, and in any case are fixed by general covariance.  Likewise, interactions with the goldstino and the transverse modes of the gravitino are dictated by supercurrent conservation in the underlying theory.  For the purpose of understanding phenomenology at colliders and in cosmology, one's main concern is to ascertain the effects of SUGRA on the vacuum structure and particle spectrum of a given SUSY model.  In this case the full machinery SUGRA can obfuscate rather than illuminate the physics.

In this paper we show SUGRA and its many complications can be dramatically simplified by applying an appropriately chosen gauge fixing or, equivalently, a prescient K\"ahler transformation.  Our starting point will be the so-called conformal compensator formalism \cite{conformalcompensator}, which is well-suited to some but not all practical calculations. In the conformal compensator formalism, one accounts for the most important SUGRA effects by augmenting the usual superspace formalism of global SUSY with a conformal compensator superfield $\bPhi$.   In the literature, the standard gauge fixing yields
\be
\label{eq:phinaive}
{\bPhi} = 1 + \theta^2 F_\Phi,
\ee
where $F_\Phi$ is the scalar auxiliary field of SUGRA.\footnote{Throughout this work, boldface and regular typeface will denote superfields and component fields, respectively.  Moreover, given a superfield $\bs X$, we will denote its lowest component by $X$.}  The conformal compensator couples to chiral superfields ${\bs X}^i$ and vector superfields ${\bs V}^a$ via the SUGRA action
\begin{align}
\label{eq:naiveconformalcompensator}
\mathcal{L}_{\rm SUGRA} &= -3 \int d^4 \theta \; \bPhi^\dagger \bPhi \; e^{-\bs{K}/3} + \int d^2 \theta \; \bPhi^3 \; {\bs W} + \mathrm{h.c.} \nonumber \\
& \qquad ~ + \frac{1}{4} \int d^2 \theta \; {\bs f}_{ab} {\bs W}^{a\alpha} {\bs W}^{b}_{\alpha} + \mathrm{h.c.} + \ldots,
\end{align}
where the K\"ahler potential $\bs K $ is a gauge invariant function of chiral and vector superfields, and the
superpotential $\bs W$ and gauge kinetic function ${\bs f}_{ab}$ are holomorphic functions of  chiral superfields.  Here, the ellipsis ($\ldots$) denotes terms involving the graviton, gravitino, and vector auxiliary field, and we work in natural units where $\MPl = 1$.

Famously, anomaly mediation \cite{anomalymediation} is most easily understood via conformal compensator methods.  More recently, this formalism has been applied to the case where multiple sequestered SUSY breaking sectors give rise to a corresponding multiplicity of goldstini \cite{goldstini1,goldstini2}.  The conformal compensator offers a simple way of understanding how goldstini obtain a universal tree-level mass of $2m_{3/2}$ \cite{goldstini1} in theories of $F$-term SUSY breaking, and illuminates modifications which can arise in certain ``goldstini variations'' \cite{goldstinivariations} and theories with imperfect sequestering \cite{pseudogoldstini}.

Despite the utility of the conformal compensator method, there are many situations where a naive application of $\bPhi$ from \Eq{eq:phinaive} leads to incorrect answers, in particular if the terms denoted by the ellipsis in \Eq{eq:naiveconformalcompensator} are improperly ignored.  For example, the ellipsis contains non-minimal couplings between matter fields and the graviton, so to properly calculate the spectrum and couplings for matter fields,
one must first Weyl rescale the metric to canonically normalize the Einstein-Hilbert action.  Likewise, in the presence of no-scale SUSY breaking, \Eq{eq:naiveconformalcompensator} simply yields the {\it wrong} fermion mass matrix unless kinetic mixing terms between matter fermions and the gravitino are properly included.  Given these complications, it is perhaps unsurprising that much of the SUGRA literature simply abandons the superfield version of SUGRA in \Eq{eq:naiveconformalcompensator} altogether and simply works in the component form.

In this work, we present an alternative gauge fixing in which \Eq{eq:naiveconformalcompensator} can be employed while entirely \emph{ignoring} additional terms involving the gravity multiplet.  Hence, the mixing terms are eliminated,  \`{a} la the $R_\xi$ gauges of spontaneously broken Yang-Mills theory.  In this gauge fixing, the conformal compensator is written as
\bea
\bPhi &=&   e^{{\bs Z}/3} (1+ \theta^2 F_{ \Phi} ),  \label{eq:phicorrect}  \\
{\bs Z} &=&  \left\langle K /2 - i \Arg W \right \rangle + \langle K_i \rangle { \bs X}^i  \label{eq:phicorrectZ},
\eea
where hereafter $\langle \rangle$ will denote a vacuum expectation value (vev) and $i$ and $\bar i$ subscripts will denote differentiation with respect to $X^i$ and $X^{\bar{i}\dagger}$, respectively.  Note that we are working in a ``zero vev'' basis in which $X^i$ has been appropriately shifted such that $\langle X^i\rangle =0$.\footnote{To work in an arbitrary vev basis, on can simply replace ${\bs X}^i$ with $({\bs X}^i - \vev{X^i})$.  The presence of terms in \Eq{eq:phicorrect} which depend explicitly on vevs may be unintuitive.  After all, the values of these vevs are unknown without computing the vacuum structure of the theory, which is in turn dependent on the vevs.  However, we will show in \Sec{subsec:vevs} that one can solve for these vevs self-consistently.}    

Given this choice of $\bs \Phi$, one recovers the correct tree-level spectrum and couplings for matter and gauge fields to leading order in $1/\MPl$ and including all effects proportional to $m_{3/2}$ without needing to perform any component manipulations.  Thus, we can effectively decouple any complications posed by the gravity multiplet from calculations involving the matter fields alone.

Alternatively, the above gauge choice can be interpreted as a well-chosen K\"ahler transformation.  In particular, we use the fact that tree-level supergravity is invariant under  
\be
\label{eq:Kahlertransformation}
\bK \rightarrow {\bs K} - {\bs Z}- {\bs Z}^\dagger, \qquad {\bW} \rightarrow e^{{\bs Z}}{\bs W },
\ee
where $\bZ$ is any chiral multiplet.   The gauge fixing in \Eq{eq:phicorrect} is equivalent to a prescient choice for ${\bs Z}$ which simply cancels the problematic terms in the K\"ahler potential which are linear in superfields, allowing one to use the standard form of $\bPhi$ from \Eq{eq:phinaive}.  Linear terms in the K\"ahler potential typically imply no-scale SUSY breaking, so this K\"ahler transformation effectively converts no-scale SUSY breaking into more familiar $F$-term breaking.  As an added bonus, in this K\"ahler basis the tree-level equation of motion for $F_\Phi$ always yields
\be
\label{eq:fphiism32}
\vev{F_\Phi} = m_{3/2}
\ee
after adjusting the cosmological constant (c.c.) to zero, so it is straightforward to identify SUGRA effects proportional to $m_{3/2}$.

Our proposed gauge fixing is related to an ``improved gauge fixing" discussed many years ago by Kugo and Uehara \cite{kugouehara}.\footnote{For a recent study of SUGRA gauge fixing in the context of inflation, see \Ref{Kallosh}.}  In that context, the improved gauge fixing was used merely as means to more efficiently calculate the component SUGRA action.  Moreover, the gauge fixing of \Ref{kugouehara} has no simple superfield realization like \Eq{eq:phicorrect}, and residual manipulations of the component SUGRA action were  required to obtain the matter spectrum and couplings.  Here, we avoid those complications at the expense of ignoring higher order $1/\MPl$ suppressed effects.

The remainder of the paper is organized as follows.  In \Sec{sec:SUGRA}, we review the formalism of conformal supergravity and show why the standard gauge fixing is suboptimal.  We then derive our preferred gauge choice in \Sec{sec:derivation}, and present a number of consistency checks.  We conclude in \Sec{sec:conclusions}, leaving further calculational details to the appendices.   In a companion paper \cite{companion}, we will use our novel gauge fixing to properly calculate the mass spectrum of goldstini and modulini in general theories of $F$-term, $D$-term, and no-scale SUSY breaking.

\section{Structure of Supergravity}
\label{sec:SUGRA}

We begin by with a brief review of conformal SUGRA, highlighting the subtleties and deficiencies of the standard gauge fixing.

\subsection{Conformal Supergravity}

As is well-known, the SUGRA action can be derived by reinterpreting minimal SUGRA as a gauge fixing of conformal SUGRA \cite{conformalcompensator}.  The purpose of the conformal compensator is to realize this gauge fixing in superspace.

The gauge redundancies of minimal SUGRA are diffeomorphisms, local Lorentz transformations, and local supersymmetry.  The additional gauge redundancies of conformal SUGRA are: local dilatations $\hat{D}$, local $U(1)_R$ chiral transformations $\hat{A}$, conformal supersymmetry $\hat{S}_\alpha$, and special conformal transformations $\hat{K}_\mu$.  The special conformal transformations can be fixed by setting the dilatation gauge field to zero,\footnote{Given the upcoming discussion in \Sec{eq:gaugefreedoms}, one might wonder whether alternative gauge fixings for $\hat{K}_\mu$ might simplify other aspects of the SUGRA action.  We were unable to find any such simplifications.} but this leaves two real gauge freedoms, $\hat{D}$ and $\hat{A}$, and a Weyl spinor gauge freedom, $\hat{S}_\alpha$.

In order to fully gauge fix conformal SUGRA to ordinary SUGRA, one introduces the conformal compensator $\bPhi$, which is a chiral superfield with conformal weight 1.\footnote{There are of course alternative choices for how to introduce compensator multiplets.  This particular choice is referred to in the literature as ``$n = -1/3$''.}  Before gauge fixing, the components of $\bPhi$ are given by
\be
\label{eq:Phicomponents}
\bPhi = \{ \sigma, \sigma \zeta_\alpha, \sigma F_\Phi \},
\ee
where the overall factor of $\sigma$ is unconventional, but convenient for later purposes.  A description of how to express these components in superspace is given in \App{app:covders}, where the main subtlety is that a multiplet with non-zero conformal weight has additional couplings to the vector auxiliary field $b_\mu$ in the gravity multiplet.  In particular, \Eq{eq:phinaive} is secretly hiding relevant terms involving $b_\mu$.

Under the dilatation and chiral transformations parametrized by a complex number $\lambda$, and the superconformal transformation parametrized by $\rho_\alpha$, the components of $\bPhi$  transform as:
\be
\label{eq:gaugefixing}
\sigma \rightarrow e^\lambda \sigma, \qquad \zeta_\alpha \rightarrow \zeta_\alpha + \rho_\alpha.
\ee 
Thus, the lowest and fermionic components of $\bPhi$ are pure gauge modes, and one can use the remaining extra gauge freedoms to set 
\be
\label{eq:stdgaugechoice}
\hat{D}: |\sigma| = 1, \qquad \hat{A}: \Arg{\sigma} = 0, \qquad \hat{S}_\alpha: \zeta_\alpha = 0.
\ee
This is the conventional gauge choice which results in \Eq{eq:phinaive}.  As we will soon see, the SUGRA action suggests a more convenient gauge fixing for practical computations.

\subsection{SUGRA Action}

To construct a valid SUGRA action, we must know the conformal weights of all fields in the theory.  Besides the conformal compensator, all other chiral superfields have conformal weight 0, and we will denote their components by
\bea
\label{eq:chiralnotation}
{\bs X}^i &=& \{X^i, \chi_\alpha ^i, F^i\}.
\eea
Vector multiplets also have conformal weight 0, but since SUSY covariant derivatives $D_\alpha$ and $\bar{D}_{\dot{\alpha}}$ have conformal weight $1/2$, the gauge field strengths ${\bW}_\alpha^a$ have conformal weight 3/2.  

In a superconformal theory, the only objects which can be consistently coupled to conformal gravity are real multiplets $\bs \Xi$ with conformal weight 2 and chiral multiplets $\bs \Sigma$ with conformal weight 3.\footnote{In some of the SUGRA literature, the $w=2$ vector multiplet $\bs \Xi$ is expressed as a $w=3$ chiral multiplet, namely the object $\bar{D}^2 \bs \Xi$ from global superspace.  We find that using $\bs \Xi$ directly is more transparent for practical calculations.}  In general, $\bs \Xi$ and $\bs \Sigma$ will be composite multiplets, and we indicate their components by
\bea
{\bs \Xi}_{(w=2)} &=& \{ C, \xi_\alpha, M, A_\mu, \lambda_\alpha, D \}, \\
{\bs \Sigma}_{(w=3)} &=& \{z, \chi_\alpha, F\}.
\eea
As argued in \App{app:covders}, once we choose the appropriate gauge fixing for $\bPhi$, one can regard $\bs \Xi$ and $\bs \Sigma$ as ordinary \emph{global} superfields (with the corresponding expansion in $\theta$ and $\bar{\theta}$) for any calculations involving matter fields alone.

From ${\bs \Xi}$ and ${\bs \Sigma}$, one can construct superconformally invariant $D$-term and $F$-terms, respectively:\footnote{These expressions differ from those in \Ref{kugouehara} because we use the two-component fermion notation of Ref.~\cite{Wess:1992cp}.  In addition, we employ different minus sign and factor of 2 conventions.}
\bea
\left[ {\bs \Xi} \right]_D &=& \frac{1}{2} e D -  \frac{1}{2} e\left( \lambda {\sigma}^\mu \overline{\psi}_\mu - i \xi \sigma^{\mu \nu} D^c_\mu \psi_\nu + \text{h.c.}\right) \nonumber \\
&& ~ + \frac{C}{3} \left( \frac{1}{2} e R - \mathcal{L}_{\rm RS}\right) + \ldots,  \label{eq:compexpD} \\
\left[{\bs \Sigma} \right]_F &=& e \left(F  - i \sqrt{2} \chi \sigma^\mu \overline{\psi}_\mu  - z \overline{\psi}_\mu \overline{\sigma}^{\mu \nu} \overline{\psi}_\nu \right), \label{eq:compexpF}
\eea
where $e$ is the determinant of the metric, $R$ is the Ricci scalar, $\mathcal{L}_{\rm RS} = \epsilon^{\mu\nu \alpha \beta} \overline{\psi}_\mu \overline{\sigma}_\nu \partial_\alpha \psi_\beta + \ldots$ is the massless Rarita-Schwinger action for the gravitino $\psi_\mu$, and $D^c_\mu = \partial_\mu + \ldots$ is the SUGRA-covariant derivative.  The ellipsis in $\left[ {\bs \Xi} \right]_D$ represents terms that are quadratic in the gravitino, but don't contribute to the gravitino mass or kinetic term.  While it is possible to rewrite these $D$-term and $F$-term invariants as superspace integrals involving the gravity supermultiplet, it is more convenient to express only the matter part of the action in (global) superspace, leaving couplings to the gravity multiplet in component form.

Since ordinary matter multiplets have vanishing conformal weight and $\bs \Xi$ ($\bs \Sigma$) has conformal weight 2 (3), the conformal compensator $\bPhi$ is necessarily present in any SUGRA action containing matter.  Given the K\"ahler potential $\bK$, superpotential $\bW$, and gauge kinetic function ${\bs f}_{ab}$, we can construct the following fully superconformally invariant action at tree-level:\footnote{With radiative corrections, one needs to account for a conformally-violating regulator, which introduces additional dependence on $\bPhi$.}
\begin{align}
\label{eq:FullSUGRA}
\mathcal{L}_{\rm SUGRA} =&~ \left[-3 \; \bPhi^\dagger \bPhi \;  e^{-{\bs K}/3}\right]_D +  \left[ \bPhi^3 \; {\bs W}\right]_F + \mathrm{h.c.} \nonumber \\ 
& ~ + \left[\frac{1}{4} {\bs f}_{ab} {\bs W}^{a \alpha} {\bs W}^{b}_{\alpha}\right]_F + \mathrm{h.c.}
\end{align}
By expanding out \Eq{eq:FullSUGRA}, we recover \Eq{eq:naiveconformalcompensator} as desired.   Note that all interactions with the gravity multiplet, as well as the graviton and and gravitino kinetic terms, come from the covariant forms of the corresponding $D$- and $F$-term expressions in \Eqs{eq:compexpD}{eq:compexpF}.  While in general the action can include additional terms involving SUSY covariant derivatives $D_\alpha$ and $\overline{D}_{\dot{\alpha}}$, we will neglect this complication in the present work.   

\subsection{Problematic Terms}
\label{sec:problematic}

The standard approach in the existing SUGRA literature is to expand \Eq{eq:FullSUGRA} in components using the definition of  $\bPhi$ in \Eq{eq:phinaive}.   However, one sees immediately from \Eqs{eq:compexpD}{eq:compexpF} that this gauge choice yields problematic terms that mix gravity and matter fields and must be carefully accounted for in any actual calculation.   These problematic terms pertain to:

\begin{list}{\labelitemi}{\leftmargin=1.5em} 
\item[{ (i)}]
\textbf{graviton normalization and kinetic mixing}:
\be
\label{eq:issue1}
\frac{C}{3} \frac{R}{2}.
\ee
At the very minimum, $\vev{C/3}$ must be set equal to $-\MPl^2$ in order to canonically normalize the Einstein-Hilbert action $R$.  In addition, if $C$ depends linearly on the matter multiplets, then there are undesirable kinetic mixing terms between the matter fields and the graviton proportional to $\vev{C_i}$ and $\vev{C_{\bar i}}$.  Note that $\vev{R} = 0$ in the flat space vacuum, so one need not worry about mass corrections arising if $C$ depends quadratically on matter fields.  Also note that if the Einstein-Hilbert term is canonically normalized, then so is the Rarita-Schwinger action.

\item[{ (ii)}]

 \textbf{gravitino kinetic mixing}:
\be
\label{eq:issue2}
i \xi \sigma^{\mu \nu} \partial_\mu \psi_\nu + \text{h.c.}
\ee
A kinetic mixing term between the gravitino and matter fields implies a non-canonical gravitino multiplet with the wrong Rarita-Schwinger action.  In contrast,
mass mixings of the form $\eta \sigma^\mu \overline{\psi}_\mu$ are perfectly healthy, since $\eta$ can be identified as the goldstino from spontaneous SUSY breaking which is eaten by the gravitino in unitary gauge.  (See \Eq{eq:goldstinocoupling}.)
\item[{ (iii)}]
 \textbf{gravitino mass phase}:
\be
\label{eq:issue3}
- z^\dagger \psi_\mu \sigma^{\mu \nu} \psi_\nu  + \text{h.c.}
\ee
The usual gravitino mass parameter is defined as a real number, so $\Arg \vev{z}$ should be set equal to zero.
\end{list}
In the standard approach, these three problem terms are eliminated by (i) performing a field-dependent Weyl rescaling of the metric, (ii) applying a shift to the gravitino which is distinct from going to unitary gauge, and (iii) performing a chiral rotation of the fermions.  While these manipulations are perfectly well-defined in terms of component fields, no simple interpretation exists in terms of superfields.   That is to say, the terms in the ellipsis  of \Eq{eq:naiveconformalcompensator} hide relevant mixing terms between the gravity multiplet and the matter multiplets.

In addition to the above three complications, there is a less apparent fourth---a marginal operator mixing the vector auxiliary field $b_\mu$ with scalars
\be
\label{eq:bad_aux_mixing}
b_\mu \partial^\mu \phi.
\ee
This term is not readily visible in \Eq{eq:compexpD}, and arises because the conformal compensator has conformal weight 1.  (See \App{app:covders}.)   Due to these interactions, integrating out the vector auxiliary field generates additional scalar kinetic terms, again mixing the gravity and matter multiplets.

\section{A Novel Gauge Fixing}
\label{sec:derivation}

Next, let us demonstrate how complications from mixing with gravitational modes can be straightforwardly removed by an appropriate choice of gauge.

\subsection{Exploiting Gauge Freedoms}
\label{eq:gaugefreedoms}

From \Eq{eq:gaugefixing}, we see that the $\hat{D}$, $\hat{A}$, and $\hat{S}_\alpha$ gauge freedoms can be spent to fix the components $\sigma$ and $\zeta_\alpha$ in $\bPhi$ equal to \emph{any} desired values.   In particular, we can even set $\sigma$ and $\zeta_\alpha$ to field-dependent functions of the matter fields, and we will exploit this freedom to manifestly resolve the three complications discussed in \Sec{sec:problematic}.

In Wess-Zumino gauge for the gauge fields, but not yet gauge fixing $\bPhi$, \Eq{eq:FullSUGRA} implies that
\begin{align}
C &= - 3 \sigma^\dagger \sigma e^{- K/3},\\
\xi_\alpha &= 3 i \sqrt{2} \sigma^\dagger \sigma e^{- K /3} \left( \zeta_\alpha - \frac{{K_i}}{3} \chi_\alpha^i \right),\\
\qquad \Arg[z] &= \Arg [\sigma^3 W].
\end{align}
Thus, we see that there is sufficient freedom in $\sigma$ and $\zeta_\alpha$ to set $C = -3$, $\xi_\alpha = 0$, and $\Arg z = 0$ to all orders in the fields.  This is essentially equivalent to the gauge choice advocated in \Ref{kugouehara}, which is reviewed in \App{app:kugoUehara}.  However in this gauge, one is forced to integrate out the vector auxiliary field of SUGRA in components, which is at odds with our aim to describe SUGRA solely in the language of superfields.

For the purposes of calculating the spectrum and couplings of matter fields, it is actually more convenient to impose a less stringent gauge choice where we only impose the conditions $C = -3$, $\xi_\alpha = 0$, and $\Arg z = 0$ to leading order in field fluctuations:
\bea
\sigma &=& \exp \left[ \frac{1}{3} \left(  \left\langle K/2 - i \Arg W \right \rangle + \langle K_i \rangle X^i \right) \right],  \label{eq:component_Phi_fix_sigma} \\ 
\zeta_\alpha &=& \frac{1}{3}\langle  K_i \rangle \chi^i_\alpha, \label{eq:component_Phi_fix}
\eea
where again we are working in a ``zero vev'' basis where $\vev{X^i} = 0$.  This gauge fixing is sufficient to ensure that there are no linear mixing terms between matter field fluctuations and the graviton/gravitino.  In addition, as long as one is working in flat space, one can ignore additional scalar mass terms from the remaining quadratic non-minimal couplings to the Einstein-Hilbert term.

\subsection{Going to Superspace}

While the gauge choice in \Eq{eq:component_Phi_fix} successfully decouples the matter fields from the gravity multiplet, it is rather inconveniently written in terms of components.  Fortunately, modulo a field redefinition of the auxiliary field $F_\Phi$, this gauge choice  can be rewritten compactly in terms of superfields as \Eq{eq:phicorrect}, repeated for convenience:
\begin{align}
\bPhi &=   e^{{\bs Z}/3} (1+ \theta^2 F_{ \Phi} ), \tag{\ref{eq:phicorrect}}\\
{\bs Z} &=  \left\langle K/2 - i\Arg W  \right \rangle + \langle K_i \rangle { \bs X}^i. \tag{\ref{eq:phicorrectZ}}
\end{align}
Were $\bPhi$ an ordinary superfield, then the mapping from \Eq{eq:component_Phi_fix} to \Eq{eq:phicorrect} would occur without complication, but since $\bPhi$ has conformal weight 1, there is an additional subtlety.

Recall that matter fields participate in additional interactions with the gravity multiplet beyond those explicitly shown in \Eqs{eq:compexpD}{eq:compexpF}.  These couplings are dictated by the full SUGRA invariance of the theory.  For example, when expanding the composite multiplets $\bs \Xi$ and $\bs \Sigma$, one encounters SUGRA-covariant derivatives $D_\mu^c$ containing additional couplings to the graviton and gravitino.  However, for calculations involving these matter fields alone, these additional  terms are irrelevant.\footnote{One might worry about additional contributions to the problematic terms in \Eq{eq:issue1},  \Eq{eq:issue2}, or  \Eq{eq:issue3}, but these are absent.}  Similarly, for matter fields of conformal weight 0, integrating out the vector auxiliary field only generates $1/\MPl^2$ suppressed dimension 6 operators, which can also be ignored.

Nevertheless, a complication still arises because the conformal compensator has conformal weight 1, and as shown in \App{app:covders}, it couples non-minimally to the vector auxiliary field.  A SUGRA-covariant derivative acting on $\sigma$ yields
\be
\label{eq:covdersigma}
D^c_\mu \sigma = \partial_\mu \sigma - \frac{i}{2} b_\mu \sigma + \ldots,
\ee
and since $\vev{\sigma} \not = 0$, linear terms involving the vector auxiliary field $b_\mu$ appear in the action.  

Fortunately, one can prove via explicit computation (see \App{app:vectoraux}) that in our gauge fixing in \Eq{eq:component_Phi_fix}, $b_\mu = 0$ to leading order in $\MPl$ and can thus be ignored.  As advertised, our gauge fixing can indeed be written in the form of \Eq{eq:phicorrect} up to irrelevant $1/\MPl$-suppressed operators.  

It is instructive to understand why $b_\mu$ can be completely ignored in our gauge but in contrast must be carefully included in the gauge proposed by the authors of \Ref{kugouehara}.  In \Ref{kugouehara}, $\sigma$ was chosen to give $C = -3$ to all orders in the field expansion, which implied that $\sigma$ was a function of both chiral and anti-chiral fields.  However, $\sigma$ itself is the lowest component of a chiral multiplet, so $D^c_\mu \sigma$ should also be a chiral object.  Since spacetime derivatives on $\sigma$ give derivatives on both chiral and anti-chiral fields, the $b_\mu$ equation of motions must cancel off the derivatives on the anti-chiral fields in order to preserve the known holomorphic structure of SUSY.  In our gauge fixing, $\sigma$ is only dependent on chiral fields since vevs are just constant complex numbers---thus holomorphicity is manifest, and $b_\mu$ can be set to $0$ at leading order.

\subsection{Understanding VEVs}
\label{subsec:vevs}

Equation (\ref{eq:phicorrect}) is the primary result of this paper, and it is worth understanding why explicit vevs are appearing in our  gauge choice.  

Two of the three problematic terms in \Sec{sec:problematic} are associated with linear terms in the K\"ahler potential.  In particular, if $K_i$ were equal to zero, then there would be no quadratic mixing terms in \Eq{eq:issue1} or \Eq{eq:issue2}.  In addition, as shown in \App{app:vectoraux}, the problematic mixing with $b_\mu$ in \Eq{eq:bad_aux_mixing} would vanish if $K_i = 0$.  For a general theory, $K_i$ will not equal zero everywhere in field space, so the best we can hope for is that $\vev{K_i} = 0$ at the minimum of the potential.  Interpreting our gauge fixing as the K\"ahler transformation in \Eq{eq:Kahlertransformation}, we can indeed remove such linear terms by the appropriate choice of $\bs Z$, but that choice will explicitly depend on $\vev{K_i}$.  In this sense, the appearance of vevs in our gauge fixing is unavoidable.  

Crucially, classical equations of motion are not affected by the appearance of vevs.  For a general function $f(x)$, the solution to $\vev{\partial f / \partial x} = 0$ is the same for $f$ as it is for the first-order Taylor expansion
\be
\tilde{f}(x) = \vev{f(x)} + \vev{f'(x)} (x - \vev{x}),
\ee
or even any linear combination of $f$ and $\tilde{f}$.  Thus, one can self-consistently solve for the $\vev{K_i}$ terms in \Eq{eq:phicorrect} by treating $\vev{K_i}$ as numbers whose values are determined by the scalar equations of motion.  Of course, $f$ and $\tilde{f}$ have different quadratic terms, and we will verify in the next subsection that scalar masses take on the expected SUGRA values.

Finally, the appearance of vevs in the gauge fixing means that \Eq{eq:phicorrect} is not manifestly (Yang-Mills) gauge invariant if there is a charged field $\bX^i$ that gets a vev.\footnote{In addition, \Eq{eq:phicorrect} is expressed in a ``zero vev'' basis where $\vev{X^i} = 0$, further obscuring (Yang-Mills) gauge invariance.}  This is not really an issue, of course, since a charged field getting a vev implies spontaneous gauge symmetry breaking.  However, one should remember that calculations using \Eq{eq:phicorrect} are only correct up to $1/\MPl^2$ suppressed dimension six operators, so there will in general be (Yang-Mills) non-invariance in the dimension six interactions.  Note that the form of \Eq{eq:phicorrect} already assumed Wess-Zumino gauge for the gauge multiplets, and one would need to redo the analysis of \Sec{eq:gaugefreedoms} to find the best SUGRA gauge fixing if one wanted to use, say, unitary gauge for the massive vector multiplets.

\subsection{Consistency Checks}
\label{sec:consistency}

To finish our discussion, we wish to show that our gauge choice satisfies a number of consistency checks.  For simplicity, we will ignore the contributions from vector multiplets for this discussion, but one can verify that gauge interactions also turn out as expected.  Since SUGRA is known to be K\"ahler invariant, the final results should be written in terms of the invariant K\"ahler potential
\be
\label{eq:defG}
G \equiv K + \log{W} + \log{W^*}.
\ee
Note, however, that K\"ahler invariance is not manifest in our gauge choice, since the gauge fixing is equivalent to picking a preferred K\"ahler basis where gravity can be decoupled.  (See \App{app:kugoUehara} for a gauge choice with manifest K\"ahler invariance but other complications.)

From \Eq{eq:compexpF} and \Eq{eq:FullSUGRA}, we see that the gravitino mass is given by
\be
\label{eq:m32check}
m_{3/2} =  \langle \left. \bPhi^3 \bW \right|_{\theta^0} \rangle = e^{\langle G \rangle / 2}.
\ee
This is the familiar K\"ahler invariant form of the gravitino mass, as desired.

If one ignores the ellipsis in \Eq{eq:naiveconformalcompensator}, then familiar global SUSY techniques can be used to derive the spectrum and couplings of matter fields.  As derived in \App{app:kinetic}, the scalar and fermion kinetic terms are proportional to the K\"ahler metric $\vev{G_{i\bar{j}}}$, leading to
\be
- \vev{G_{i \bar{j}}} \partial_\mu X^i \partial^\mu X^{\dagger\bar{j}} ,
\ee
and
\be
- \vev{G_{i\bar{j}}} \overline{\chi}^{\bar{j}} i \overline{\sigma}^\mu \partial_\mu \chi^i,
\ee
for the scalars and the fermions, respectively.  Note the appearance of vevs in these expressions, as higher order field couplings to the kinetic terms differ from the exact SUGRA results at order $1/\MPl^2$.

The scalar potential is derived in \App{app:scalar}. After solving the $F_\Phi$ and $F^i$ equations of motion, we find
\begin{equation}
V = e^G  \left(G^{i} G_{i} - 3\right) e^{-2K_{\text{quad}}/3},
\label{eq:ourV}
\end{equation}
which depends on the K\"ahler potential starting at quadratic order in the fields
\be
\label{eq:Kquad}
K_{\text{quad}} \equiv K - \langle K \rangle -  \langle K_i \rangle X^i - \langle K_{\bar{j}} \rangle X^{\dagger \bar{j}}.
\ee
In \Eq{eq:ourV}, indices are raised and lowered with the inverse metric $G^{i\bar{j}}$ (without a vev), such that $G^i \equiv G^{i\bar{j}}G_{\bar{j}}$.  

We see that the condition for vanishing c.c.\ is the same as in exact SUGRA
\be
\vev{G_i G^i} = 3, 
\ee
so the vacuum structure is maintained.  After tuning the c.c.\ to zero, the equation of motion for $F_\Phi$ yields
\be
\vev{F_\phi} = m_{3/2} ,
\ee
as advertised in \Eq{eq:fphiism32}.  Note that \Eq{eq:ourV} matches the SUGRA scalar potential to leading order in $1/ \MPl$, and includes all the effects proportional to $m_{3/2}$.  In particular, the $K_{\text{quad}}$ term would appear to give corrections at quadratic order in fields, but these vanish once the c.c. is tuned to zero.  

Most important for our companion paper \cite{companion}, we recover the SUGRA results for the fermion spectrum.  Before going to unitary gauge for the gravitino, the (normalized) goldstino mode $\eta_{\rm eaten}$ couples the gravitino as
\be
\label{eq:goldstinocoupling}
- i \frac{m_{3/2}\sqrt{3}}{\sqrt{2}}  \eta_{\rm eaten} \sigma^\mu \overline{\psi}_\mu + \text{h.c.}
\ee
From \Eqs{eq:compexpD}{eq:compexpF} and \Eq{eq:FullSUGRA}, and using the $F^i$ equation of motion, we can identify the goldstino mode as
\be
\label{eq:goldstinomode}
\eta_{\rm eaten} = \frac{1}{\sqrt{3}} \vev{G_i} \chi^i.
\ee
As shown in \App{app:fermion}, the complete fermion spectrum (including the goldstino) is
\be
-\frac{1}{2}m_{ij} \chi^i \chi^j + \mathrm{h.c.}, \quad m_{ij} = m_{3/2} \left \langle \nabla_i G_j + G_i  G_j \right \rangle ,
\ee
where $\nabla_i G_j \equiv \partial_i G_j - \Gamma^k_{ij} G_j$ depends on the Christoffel symbol $\Gamma^k_{ij}$ derived from the K\"{a}hler metric.  This mass matrix is not particularly illuminating in and of itself, but the fact that it can be derived entirely within the superspace formalism will allow us to easily compute the spectrum of fermion masses in a companion paper \cite{companion} on goldstini from $F$-term, $D$-term, and no-scale SUSY breaking.

\section{Conclusions}
\label{sec:conclusions}

In this paper we have derived a novel gauge choice for conformal SUGRA which results in a simplified version of the tree-level SUGRA Lagrangian.   Our improved gauge  is manifestly better suited for phenomenological applications, and moreover  is an improvement over the gauge choice of \Ref{kugouehara} since it decouples matter modes not only from the graviton and gravitino but also from the vector auxiliary field.  Hence, SUGRA calculations involving matter fields alone can be performed directly in (global) superspace, including effects proportional to $m_{3/2}$.  

This gauge fixing can also be understood as the presciently chosen K\"ahler transformation in \Eq{eq:Kahlertransformation}.  This is not surprising, since a K\"ahler transformation
\be
\tag{\ref{eq:Kahlertransformation}}
\bK \rightarrow {\bs K} - {\bs Z}- {\bs Z}^\dagger, \qquad {\bW} \rightarrow e^{{\bs Z}}{\bs W }
\ee 
has the same effect as a gauge transformation on the conformal compensator (plus a field redefinition of $F_\phi$)
\be
\bPhi \rightarrow e^{\bZ/3} \bPhi.
\ee
In this paper, we have exploited this K\"ahler redundancy to simplify SUGRA calculations.  

This formulation makes certain aspects of  SUGRA  manifest while obscuring others.  By expanding \Eq{eq:FullSUGRA} as \Eq{eq:naiveconformalcompensator}, we have emphasized the spectrum and couplings of matter fields but have hidden gravitational interactions.  On the other hand, the couplings of the graviton and gravitino are given at leading order by the well-known stress-energy tensor and supercurrent, respectively, so little is lost by hiding them.  Crucially, \Eq{eq:naiveconformalcompensator} still includes all couplings to the goldstino degree of freedom identified in \Eq{eq:goldstinomode}, which is phenomenologically more relevant than the transverse gravitino modes in the goldstino equivalence limit.

In a companion paper \cite{companion}, we will use this novel gauge fixing to calculate the spectrum of goldstini in general theories of $F$-term, $D$-term, and ``almost no-scale'' SUSY breaking.  With standard component SUGRA methods, such calculations would be tedious and obscure, but in superspace, they become straightforward and transparent.  Obviously, one would like to generalize this gauge fixing procedure beyond minimal SUGRA models, by properly including radiative corrections and SUSY-covariant derivatives in the action.  Ultimately, one hopes that this alternative gauge fixing will shed light on the more general properties of SUGRA beyond phenomenology.

\begin{acknowledgments}
We thank Allan Adams, Martin Ro\v{c}ek, and Matt Strassler for helpful conversations.  C.C is supported in part by the Director, Office of Science, Office of High Energy and Nuclear Physics, of the US Department of Energy under Contract DE-AC02-05CH11231 and by the National Science Foundation on grant PHY-0457315.  F.D. and J.T. are supported by the U.S. Department of Energy under cooperative research agreement Contract Number DE-FG02-05ER41360.  
\end{acknowledgments}

\appendix

\section{Superfields in SUGRA}
\label{app:covders}

In order to describe matter multiplets in terms of the ordinary global superspace variables $\theta$ and $\bar{\theta}$, we need to know how to package the components of a multiplet into $\theta$-dependent superfields, including any relevant SUGRA effects.  Here, we follow the logic of  \Ref{kugouehara,Kugo:1982cu}, though we use two-component fermion notation.

For a chiral multiplet $\bX$ with components
\be
{\bs X} = \{X, \chi_\alpha, F\},
\ee
we can construct a familiar looking superfield in terms of the usual global superspace variables as
\be
\begin{array}{lllllllllllllll}
{\bs X} &=& X &+& \sqrt{2} \theta \chi &+& \theta^2 F\\
&& &+& i \theta \sigma^\mu \bar{\theta} D^c_\mu X &-& \frac{i}{\sqrt{2}} \theta \theta D^c_\mu \chi \sigma^\mu \bar{\theta}   \\
&& & & &+& \frac{1}{4} \theta^4 D^c_\mu D^{c\mu} X.
\end{array}
\ee
Compared to the expressions from global SUSY, we have simply replaced the ordinary derivative $\partial_\mu$ with the SUGRA-covariant derivative $D^c_\mu$.  

The SUGRA-covariant derivative depends on the graviton, gravitino, and vector auxiliary field, with full expressions given in \Ref{Kugo:1982cu}.  For our purposes, we are  mainly interested in anomalous couplings to the vector auxiliary field $b_\mu$, which arise because the conformal compensator has conformal weight 1. The SUGRA-covariant derivative for a chiral multiplet of conformal weight $w$ is
\begin{align}
D_\mu^c X &= \left(\partial_\mu - i \frac{w}{2} b_\mu \right) X + \ldots,\\
D_\mu^c \chi_\alpha &= \left(\partial_\mu  - i \left(\frac{3}{4} - \frac{w}{2} \right) b_\mu \right) \chi_\alpha + \ldots, \label{eq:sugracovder}
\end{align}
where the ellipsis indicates additional terms involving the graviton and gravitino.

Immediately, we see that the expression in \Eq{eq:phinaive} for the na\"ive conformal compensator is incomplete.  Since the conformal compensator has conformal weight $1$, there should be additional terms involving $b_\mu$ in the $\theta \bar{\theta}$ and $\theta^4$ components.  As discussed more in \App{app:vectoraux}, we can avoid that complication by ensuring that the equations of motion fix $b_\mu = 0$ to leading order in $1/\MPl$.

We also see that for ordinary chiral multiplets with $w = 0$, there are no complications posed by $D_\mu^c$.  While there are residual couplings of $b_\mu$ to fermions, they only generate $1/\MPl^2$ suppressed dimension 6 operators.  Similarly, it is straightforward to show that the graviton and gravitino terms elided in \Eq{eq:sugracovder} do not introduce any of the problematic mixing terms discussed in \Sec{sec:problematic}.  A similar analysis holds for vector multiplets, with the same conclusions.

Thus, for calculations involving matter fields alone, we are free to use ordinary \emph{global} superfields for calculational purposes, as long as we ensure that $b_\mu = 0$ to leading order in $1/\MPl$.  The condition $b_\mu = 0$ is a key feature of the gauge choice in \Eq{eq:phicorrect}, a feature not present in \Ref{kugouehara}, as discussed in the next appendix.

\section{The Kugo-Uehara Gauge}
\label{app:kugoUehara}

The gauge choice in this paper is closely related to a gauge choice presented by Kugo and Uehara in \Ref{kugouehara}, which we review in this appendix.  The Kugo-Uehara gauge choice starts with essentially the same logic as in \Eq{eq:gaugefreedoms}, but differs from our gauge fixing by setting
\be
C = -3, \qquad \xi_\alpha = 0, \qquad \Arg z = 0, 
\ee
to all orders in the fields.  This results in the conformal compensator having components
\be
\label{eq:KUgauge}
\bPhi = \exp \left[ \frac{1}{3} \left( K/2 -i \Arg W\right) \right] \times \left\{1,  \frac{ K_i \chi^i}{3}, F_\phi \right\},
\ee
where $K$ and $W$ include their full field dependence.  By doing a field redefinition of $F_\Phi$, this is more conveniently expressed in term of the superpotential $\bW$ and invariant K\"ahler potential $G$ from \Eq{eq:defG} as\footnote{In \Ref{kugouehara}, the conformal compensator was first rescaled by the superpotential $\bPhi = \bW^{-1/3} \widetilde{\bPhi}$ and then the gauge fixing was applied to $\widetilde{\bPhi}$, obtaining the same result.}
\be
\label{eq:kugoUehara}
\bPhi = \bW^{-1/3} \widetilde{\bPhi}, \qquad \widetilde{\bPhi} = e^{G/6} \times \left\{1,  \frac{ G_i \chi^i}{3}, F_\phi \right\}.
\ee
Here, $\bW$ is a full chiral superfield, but $G$ is only a function of the scalar fields.

This gauge choice has a number of nice features.  First, this gauge choice makes K\"ahler invariance manifest.  Looking at \Eq{eq:naiveconformalcompensator}, we see that the ${\bs W}^{-1/3}$ term in $\bPhi$ precisely cancels against the superpotential $\bW$ in the action, and the very same ${\bW}^{-1/3}$ term combines with $e^{-\bK/3}$ to yield
\be
({\bs W}{\bs W}^*)^{-1/3} e^{-\bK/3} \equiv -3 e^{-{\bs G}/3}.
\ee
Thus, the SUGRA action depends only on $\bs G$, which can be thought of as the invariant K\"ahler potential lifted into superspace.  Second, this gauge choice eliminates the problematic mixing terms everywhere in moduli space, such that no matter what the field vevs are, none of the three terms from \Sec{sec:problematic} ever arise.  Finally, the fermionic component of $\widetilde{\bPhi}$ is proportional to the eaten goldstino $G_i \chi^i / \sqrt{3}$, and setting this component to zero automatically results in unitary gauge for the gravitino.

Despite the apparent simplicity of this Kugo-Uehara gauge, it suffers from a hidden problem, namely problematic mixing with the vector auxiliary field $b_\mu$ described in \Eq{eq:bad_aux_mixing}.  This means that $\bPhi$ cannot simply be lifted into global superspace, since its components have residual dependence on the SUGRA-covariant derivatives from \Eq{eq:sugracovder}.  Integrating out $b_\mu$ gives additional terms in the action which spoil the simplicity of \Eq{eq:naiveconformalcompensator}.  In addition, $\bPhi$ does not have the expected holomorphy properties of a usual chiral multiplet, since $\widetilde{\bPhi}$ depends on the full K\"ahler potential $G$, which is a function of both chiral and antichiral fields.

Thus, we prefer \Eq{eq:phicorrect} for practical calculations.  That said, the gauge choice in \Eq{eq:kugoUehara} is convenient for zero momentum calculations where the problematic term in \Eq{eq:bad_aux_mixing} is irrelevant.  Alternatively, if one wanted to keep track of the most important effects of $b_\mu$ using a superfield language, one could introduce a new auxiliary field multiplet
\be
\bs B = - \frac{1}{2}\theta \sigma^\mu \bar{\theta} b_\mu,
\ee
and make the replacement
\be
\bPhi^\dagger \bPhi \rightarrow \bPhi^\dagger e^{\bs B} \bPhi
\ee
in \Eq{eq:naiveconformalcompensator}.  Interpreting $b_\mu$ as the gauge field for local $U(1)_R$ chiral transformations, we see that $\bPhi$ indeed has conformal weight 1.\footnote{More precisely, a chiral multiplet with conformal weight $w$ has charge $\{w,w\}$ under $\hat{D}$ and $\hat{A}$, while an anti-chiral multiplet with weight $w$ has charge $\{w,-w\}$.  Recall, though, that we have used the $\hat{K}_\mu$ gauge freedom to eliminate the dilatation gauge field.}

\section{The Vector Auxiliary Field}
\label{app:vectoraux}

As described in \App{app:covders}, in order to convert global superfields into SUGRA superfields, all spacetime derivatives $\partial_\mu$ must be replaced by covariant derivatives $D^c_\mu$ which are covariant under the complete set of conformal SUGRA gauge redundancies.   This covariant derivative $D^c_\mu$ contains couplings to the gravity multiplet, which are largely irrelevant for our phenomenological purposes.  The one exception is covariant derivatives acting on the conformal compensator itself, since the compensator has conformal weight 1.  

Since we have parametrized $\bPhi$ in \Eq{eq:Phicomponents} with an overall factor of $\sigma$, we can treat $\sigma$ as the only field with conformal weight 1, with $\zeta_\alpha$ and $F_\Phi$ having the usual conformal weight 0.  Thus, the only non-trivial covariant derivative is
\be
\label{eq:covdersigmagenw}
D^c_\mu \sigma = \partial_\mu \sigma - \frac{i}{2} w b_\mu \sigma + \ldots,
\ee
where the conformal weight $w=1$ for $\sigma$, $b_\mu$ is the vector auxiliary field, and we have elided the graviton and gravitino terms for simplicity.

We can expand out \Eq{eq:naiveconformalcompensator} to isolate terms depending on the vector auxiliary field: 
\be
\label{eq:bmuterms}
3 \sigma^\dagger \sigma e^{-K/3}  \left(  \frac{b^\mu b_\mu}{4} + b^\mu \Im \left( \frac{1}{3}K_i \partial_\mu x^i -  \frac{\partial_\mu \sigma}{\sigma} \right) \right),
\ee
where we have elided additional terms that are quadratic in fermion fields.  Using the gauge fixing from \Eq{eq:component_Phi_fix_sigma}, we see that
\be
\label{eq:bmucancel}
\frac{\partial_\mu \sigma}{\sigma} = \frac{1}{3}\vev{K_i} \partial_\mu x^i
\ee
so there are no kinetic mixing terms of the form of \Eq{eq:bad_aux_mixing}, only higher order terms in the scalar field expansion.  Since $b_\mu$ has a mass of order $\MPl^2$, the only effect of the auxiliary fields is to generate $1/\MPl^2$ suppressed dimension six operators involving matter fields, which are irrelevant for our phenomenological purposes.  Therefore we are free to set $b_\mu = 0$ to leading order for this gauge choice.

It is now clear why the Kugo-Uehara gauge choice in \Eq{eq:KUgauge} has residual dependence on the vector auxiliary field.  In that gauge, $\sigma$ is a function of both chiral and anti-chiral fields, so the leading order cancellation seen in \Eq{eq:bmucancel} does not persist.  In this sense, our gauge choice is unique, since it is the minimal (local) gauge fixing that eliminates \Eq{eq:bad_aux_mixing}.  

\section{Scalar and Fermion Kinetic Terms}
\label{app:kinetic}
Using \Eq{eq:naiveconformalcompensator}, it is straightforward to check that our gauge fixing results in the expected scalar and fermion kinetic terms in SUGRA.  In particular, \Eq{eq:naiveconformalcompensator} can be expanded using standard superspace methods, up to corrections at order $1/\MPl$.

The kinetic operators can only come from the first term in \Eq{eq:naiveconformalcompensator}, and we parameterize the chiral supermultiplets as in \Eq{eq:chiralnotation}. To simplify the notation, it is convenient to introduce the superfield
\be
\bT = - 3 \bPhi^{\dag} \bPhi \, e^{- {\bs K} /3} ,
\label{eq:T}
\ee
resulting immediately in the kinetic terms 
\be
\label{eq:appscalarkin}
- T_{i \bar{j}}  \, \partial_{\mu} X^i \, \partial^{\mu}  X^{\dagger \bar{j}} \ ,
\ee
and
\be
\label{eq:appfermionkin}
- T_{i \bar{j}}  \, \overline{\chi}^{\bar{j}} i \overline{\sigma}^\mu \partial_\mu \chi^i \ ,
\ee
for the scalars and the fermions, respectively.

Using the gauge fixing for $\bPhi$ in \Eq{eq:phicorrect}, it is straightforward to find $T_{i \bar{j}}$ from \Eq{eq:T}.  Since $\bPhi$ is itself expressed as a function of the chiral multiplets, there is no complications in taking field derivatives.  We find
\bea
T &=& -3 e^{-K_{\text{quad}} / 3},\\
T_i &=& \delta K_i \, e^{-K_{\text{quad}} / 3}, \\
T_{i \bar{j}}  &=& \left(G_{i \bar{j}}  - \frac{ \delta K_i \, \delta K_{\bar{j}}}{3} \right)  e^{-K_{\text{quad}}/3},
\eea
where $\delta K_i \equiv K_i - \langle K_i \rangle$ and $K_{\text{quad}}$ is defined in \Eq{eq:Kquad}.

Because of the relationship between the vevs
\be
\vev{T_{i \bar{j}}} = \vev{G_{i \bar{j}}},
\ee
we indeed recover the correct SUGRA kinetic terms.  At higher orders in the field expansion of \Eq{eq:appscalarkin} and \Eq{eq:appfermionkin}, there will be deviations from the SUGRA predictions at order $1/\MPl^2$.

\section{Scalar Potential and Auxiliary VEVs}
\label{app:scalar}

Another check of the gauge fixing is to make sure that the vacuum structure of the theory matches the exact results from SUGRA.  Again, we can use superspace methods to analyze \Eq{eq:naiveconformalcompensator}.  It is convenient to define the superfield
\be
\bP = \bPhi^3 \bW  ,
\label{eq:P}
\ee
and we will continue to use the notation $\bT$ from \Eq{eq:T}.

The scalar potential derived from \Eq{eq:naiveconformalcompensator} is
\bea
- V &=& T_{i \bar{j}} F^i F^{\dagger \bar{j}} + T_i  F^i  F^\dagger_\Phi + T_{\bar{j}} F^{\dagger \bar{j}}  F_{\Phi} + T F_\Phi F^\dagger_\Phi  \nonumber \\ 
&&~+ P_i F^i + 3 F_{\Phi} P + P_{\bar{j}}^\dagger F^{\dagger \bar{j}} + 3 F^\dagger_{\Phi} P^\dagger. \label{eq:checkpotential}
\eea
The expression for $F_\Phi$ is obtained from its equation of motion
\be
F_\Phi = - \frac{T_i F^i + 3 P^\dagger}{T}.
\ee
Our gauge choice has $\vev{T_i} = 0$ and $\vev{T} = -3$ by construction, and $\vev{P} = m_{3/2}$ from \Eq{eq:m32check}, so
\be
\vev{F_\Phi} = m_{3/2}
\ee
as advertised.  

Substituting $F_\Phi$ back into \Eq{eq:checkpotential}, we have
\bea
- V &=& \left(T_{i \bar{j}} - \frac{T_i T_{\bar{j}}}{T}\right) F^i F^{\dagger \bar{j}} + \left( P_i - 3 P \frac{T_i}{T}\right) F^i \nonumber\\
&&~+ \left(P^\dagger_{\bar{j}} - 3 P^\dagger  \frac{T_{\bar{j}}}{T} \right) F^{\dagger \bar{j}} - 9 \frac{P P^\dagger}{T} .
\eea
Using the fact that
\bea
T_{i \bar{j}} - \frac{T_i T_{\bar{j}}}{T} &=& G_{i \bar{j}} \, e^{-K_{\text{quad}}/3} ,\\
P_i - 3 P \frac{T_i}{T} &=& G_i P,\\
P P^\dagger & = & e^G e^{-K_{\text{quad}}},
\eea
we can simplify the potential and solve for $F^i$
\be
F^i = - P^\dagger G^i \, e^{K_{\text{quad}}/3}.
\ee
Here $G^i \equiv G^{i\bar{j}}G_{\bar{j}}$ is defined in terms of the inverse K\"ahler metric.  The final expression for the scalar potential is
\begin{equation}
V = e^G  \left(G^{i} G_{i} - 3\right) e^{-2K_{\text{quad}}/3},
\end{equation}
which agrees with the SUGRA scalar potential at leading order in $ 1/ \MPl^2$ when expanded around flat space.

\section{Goldstino Mode and Fermion Spectrum}
\label{app:fermion}

The final check of our gauge fixing is to verify that the fermion structure matches the SUGRA expectation.  For simplicity, we will ignore gauginos for this discussion, and we will again use the notation of $\bT$ from \Eq{eq:T} and  $\bP$ from \Eq{eq:P}.  

The goldstino mode couples to the gravitino as in \Eq{eq:goldstinocoupling}.  Using \Eqs{eq:compexpD}{eq:compexpF}, we can identify the goldstino direction $\eta_{\rm eaten}$ as
\be
\eta_{\rm eaten} =\frac{\sqrt{2}}{ \sqrt{3} m_{3/2}}  \left(\frac{1}{2}\bT \vert_{\bar{\theta}\bar{\theta} \theta} + \bP \vert_{\theta} \right). \\
\ee
Focusing only on the minimum of the scalar potential, and using the auxiliary field equations of motion, we find
\bea
\eta_{\rm eaten} &=&  \frac{1}{\sqrt{3} m_{3/2}}\left\langle \left(F_\Phi^\dagger T_i + T_{i\bar{j}} F^{\dagger \bar{j}} \right) + 2 P_i \right\rangle \chi^i \nonumber\\
& =& \frac{1}{\sqrt{3}} \vev{G_i} \chi^i,
\eea
as desired.  With $D$-terms turned on, $\eta_{\rm eaten}$ will pick up an additional contribution from the gaugino in $\bT \vert_{\bar{\theta}\bar{\theta} \theta}$ as well as $\bW^\alpha \bW_\alpha \vert_{\theta}$. 

To check the fermion masses, we expand out \Eq{eq:naiveconformalcompensator}, looking for the operators $\chi^i \chi^j$ in the Lagrangian:
\be
- \frac{1}{2}\left \langle T_{ij\bar{k}} F^{\dagger \bar{k}} + T_{ij} F^\dagger_\Phi + P_{ij} \right \rangle \chi^i \chi^j  + \hc,
\ee
where for again we have only considered the vevs of these expressions.  Using the definitions of $T$ and $P$, we can extract
\bea
\vev{T_{ij\bar{k}}F^{\dagger \bar{k}} } &=& - m_{3/2} \vev{G_{ij\bar{k}}G^{\bar{k}}},\\
\vev{P_{ij} + T_{ij} F^\dagger_\Phi} & = & m_{3/2} \vev{G_{ij} + G_i G_j}.
\eea
Thus, the fermion mass matrix is 
\be
\mathcal{L}_{{\rm mass}} = - \frac{1}{2} m_{3/2} \left\langle \nabla_i G_j  +   G_i   G_j \right\rangle \chi^i \chi^j + \hc,  
\ee
where $\nabla_i G_j \equiv \partial_i G_j - \Gamma^k_{ij} G_k$, and the Christoffel symbol $\Gamma^k_{ij}$ is derived from the K\"{a}hler metric $G_{i j \bar{k}} = G_{m \bar{k}} \Gamma^{m}_{i j}$.  This is the expected SUGRA result.

\end{document}